# Evolution of spin excitations into the superconducting state in $FeTe_{1-x}Se_x$

M. D. Lumsden[1], A. D. Christianson[1], E. A. Goremychkin[2,3], S. E. Nagler[1], H. A. Mook[1], M. B. Stone[1], D. L. Abernathy[1], T. Guidi[3], G. J. MacDougall[1], C. de la Cruz[4], A. S. Sefat[1], M. A. McGuire[1], B. C. Sales[1], & D. Mandrus[1]

*[1]Oak Ridge National Laboratory, Oak Ridge, Tennessee 37831, USA. [2]Argonne National Laboratory, Argonne, Illinois 60439, USA. [3]ISIS Facility, Rutherford Appleton Laboratory, Chilton, Didcot OX11 0QX, UK. [4]The University of Tennessee, Knoxville, Tennessee 37996, USA.*

**The nature of the superconducting state in the recently discovered Fe-based superconductors[1-3] is the subject of intense scrutiny. Neutron scattering investigations have already elucidated a strong correlation between magnetism and superconductivity in the form of a spin resonance in the magnetic excitation spectrum[4-7]. A central unanswered question concerns the nature of the normal state spin fluctuations which may be responsible for the pairing mechanism. Here we show inelastic neutron scattering measurements of $Fe_{1.04}Te_{0.73}Se_{0.27}$, not superconducting in bulk, and $FeTe_{0.51}Se_{0.49}$, a bulk superconductor. These measurements demonstrate that the spin fluctuation spectrum is dominated by two-dimensional incommensurate excitations near the (1/2,1/2) (square lattice (π,0)) wavevector , the wavevector of interest in other Fe-based superconductors, that extend to energies at least as high as 300 meV. Most importantly, the spin excitations in $Fe_{1+y}Te_{1-x}Se_x$ exhibit four-fold symmetry about the (1,0) (square lattice (π,π)) wavevector and are described by the identical wavevector as the normal state spin excitations in the high-$T_C$ cuprates[8-12] demonstrating a commonality between the magnetism in these classes of materials which perhaps extends to a common origin for superconductivity.**

The discovery of superconductivity in $LaFeAsO_{1-x}F_x$ with $T_C$ = 28 K[1] sparked a flurry of scientific activity and $T_C$ rapidly increased to ~55 K on replacing La with other rare earth elements[2,12-14]. In addition to the RFeAsO family of compounds, superconductivity was also discovered in the $AFe_2As_2$[3], LiFeAs[15], and in the alpha phase of $Fe_{1+y}Te_{1-x}Se_x$[16,17]. These materials share common structural square layers with Fe coordinated with either a pnictogen or a chalcogen. The unit cell contains two Fe atoms generating a reciprocal space rotated by 45 degrees from the conventional square lattice (see Fig. 1a). In RFeAsO and $AFe_2As_2$, the parent compounds exhibit long-range spin density wave order characterized by the wavevector **Q** = (1/2,1/2,L) [18-20] Doping suppresses magnetic order allowing superconductivity to emerge with the concomitant appearance of a resonance in the spin fluctuation spectrum[4-7]. However, the resonance likely contains only a small fraction of the total magnetic spectral weight and, consequently, understanding the role magnetism plays in the superconductivity of these materials requires detailed understanding of the higher energy spectrum of magnetic excitations.

The $Fe_{1+y}Te_{1-x}Se_x$ materials are ideal candidates for such a study of the magnetic excitations as large single crystals, necessary for detailed inelastic neutron scattering studies, may be grown. However, these materials differ somewhat from other Fe-based superconductors in that the $Fe_{1+y}Te$ endpoint member orders magnetically with a structure described by the wavevector (1/2,0,1/2)[21] as opposed to (1/2,1/2, L). Despite this ordering wavevector, superconducting samples of $Fe_{1+y}Te_{1-x}Se_x$ with higher Se content exhibit a magnetic resonance at the same (1/2,1/2) wavevector as other Fe-based superconductors[7,22] suggesting commonality in the magnetic response. To explore the magnetic excitations, inelastic neutron scattering measurements were performed using the MERLIN spectrometer at the ISIS neutron scattering facility (x=0.27) and the ARCS spectrometer at the Spallation Neutron Source (x=0.49). Measurements of lower energy excitations were performed using the HB1 (x=0.27) and HB3 (x=0.49) triple-axis spectrometers at the High Flux Isotope Reactor. The single crystal samples of $Fe_{1.04}Te_{0.75}Se_{0.25}$ and $FeTe_{0.51}Se_{0.49}$ studied here were prepared as in ref. 23. Bulk measurements indicate weak, likely filamentary, superconductivity in $Fe_{1.04}Te_{0.73}Se_{0.27}$ and bulk superconductivity in $FeTe_{0.51}Se_{0.49}$.

Figure 2a-h summarizes the measured magnetic excitation spectrum for several energy transfers for both the x=0.27 and x=0.49 samples. Before proceeding to describe the data in detail we first note that the observed spectrum of magnetic excitations is very two-dimensional (2d) in nature. For these measurements, wavevector transfer perpendicular to the (HK0) plane varies with incident energy; consistency of the data for differing incident energies is evidence for this two-dimensionality. Furthermore, fits to a 2d model (equation 1 below) can also reproduce the measured spectrum at different sample rotations (see supplementary information) providing quantitative evidence for two-dimensionality. This is consistent with recent measurements of the magnetic resonance in a single crystal of $FeSe_{0.4}Te_{0.6}$ indicating two-dimensional excitations[22].

The low energy magnetic response (figure 2a-b) for the x=0.27 sample, is characterized by two peaks at incommensurate wavevectors near (1/2,1/2). Interestingly the data does not show four-fold symmetry around this wavevector as expected for a tetragonal system, but rather form a quartet around the (1,0) wavevector. With increasing energy, the peaks disperse away from (1/2,1/2) towards (1,0) as shown schematically in Fig. 1b. At higher energies, the excitations continue to disperse towards (1,0) but also evolve from spots into rings (Fig. 2c) centred on this wavevector. Eventually, as shown for an energy transfer of 120 meV in Fig. 2d, the excitations evolve into broad spots centred at (1,0). For the superconducting x=0.49 sample, the low energy spectrum appears as a series of asymmetric spots (Fig. 2e). However, data measured at a higher energy transfer of 22 meV (Fig. 2f) shows incommensurate peaks similar to those in the x=0.27 sample. This can easily be understood by considering that the size of the incommensuration away from (1,0) is larger in the x=0.49 sample such that the pair of peaks around (1/2,1/2) have moved closer together and overlap significantly. At an energy of 45 meV (Fig. 2g) the magnitude of the incommensuration appears similar in the two samples but the x=0.49 scattering appears to have not fully evolved into the rings of scattering present in the x=0.27 sample (Fig. 2c). At high energies, the scattering is similar in the two samples as can be seen by comparing Fig. 2d and 2h. For both compositions, the excitations persist for energy transfers as large as 300 meV with Q-dependence similar to that shown at 120 meV for all higher energies.

Examination of the wavevector describing these excitations reveals similarities with the high-$T_C$ cuprates. The quartet of peaks are characterized by wavevectors ($1\pm\xi$, $\pm\xi$) and ($1\pm\xi$, $\mp\xi$) which, in square lattice notation, corresponds to ($\pi\pm\xi,\pi$) and ($\pi$, $\pi\pm\xi$) as shown in Fig. 1b. This is precisely the same wavevector as the low energy excitations observed in $La_{2-x}Sr_xCuO_4$[8,9] and $YBa_2Cu_3O_{6+x}$[10,11] indicating remarkable commonality in the excitation spectrum of these two classes of high-$T_C$ superconductors. Furthermore, the evolution of the scattering from well defined peaks at low energies to broadened rings at higher energies is a characteristic property of magnetic excitations in the cuprates[24,25]. The magnitude of $\xi$, however, is much larger in $Fe_{1+y}Te_{1-x}Se_x$ resulting in low energy excitations displaced away from (1,0) and much closer to (1/2,1/2).

At low energies, the largest difference between the two concentrations becomes evident as shown in Fig. 3a-b (E=6 meV). In addition to the incommensurate excitations, an additional component centered near (1/2,0) is present in the x=0.27 sample (also visible in Fig. 2a). With decreasing energy, the intensity of this component increases and eventually forms the short range order observed previously for samples with a similar concentration[21,26]. It has been suggested[27] that excess Fe in Te rich samples results in local moments that may provide a pair breaking mechanism which destroys superconductivity. The component of scattering observed near (1/2,0) is absent in the x=0.49 sample with no excess Fe (Fig. 3b). Furthermore, the scattering near (1/2,0) exists inelastically for all energies below ~10 meV and, as such, exists well below the superconducting gap potentially providing a pair breaking mechanism. These observations are consistent with the additional component near (1/2,0) existing as a result of the influence of additional Fe in the x=0.27 sample.

To quantify the dispersion of the incommensurate excitations, the data was fit using the phenomenological Sato-Maki (SM) function[28] previously used for the cuprates[25]

$$\chi''(\mathbf{Q},\omega) = \chi_0(\omega)\frac{\kappa^4(\omega)}{\left[\kappa(\omega)^2 + R(\mathbf{Q})\right]^2} \tag{1}$$

where

$$R(\mathbf{Q}) = \frac{\left[(\mathrm{H}-\mathrm{H_C})^2 + (\mathrm{K}-\mathrm{K_C})^2 - \delta^2\right]^2 + \frac{\lambda}{4}\left[(\mathrm{H}-\mathrm{H_C})^2 - (\mathrm{K}-\mathrm{K_C})^2\right]^2}{4\delta^2}. \tag{2}$$

The specific form for $R(\mathbf{Q})$ in equation 2 describes the incommensurate excitations where ($H_C$,$K_C$) represents the wavevector about which the excitations are four-fold symmetric, in this case ($H_C$,$K_C$)=(1,0). As discussed below, δ parameterizes the dispersion, λ defines the evolution of the spectrum from peaks (large λ) to rings (small λ), and κ is a broadening parameter.

The result of fits to the data presented in Fig. 2a-h using the SM function convolved with instrumental resolution are shown in Fig. 2i-l(2m-p) for x=0.27(0.49). The fits agree well with the measurements over the full range of measured energies. The quality of the fits using the SM function is also demonstrated in Fig. 4d-g for cuts along the line (1/2±ξ, 1/2∓ξ) indicating excellent agreement over a wide range of wavevector and energy transfer. The same SM function can be used to describe the additional component in the x=0.27 sample near (1/2,0) with the $R(\mathbf{Q})$ factor (equation 2) rotated by 45 degrees (see supplementary information for the equation describing both components) and the fits displayed in Fig. 3c as well as Fig. 2i contain both components again yielding excellent agreement with the data.

The best fit value of $\delta$ parameterizes the dispersion of the incommensurate excitation. If we define the wavevector of the incommensurate excitation as $\mathbf{q_{inc}} = (1\pm\xi, \mp\xi)$, the incommensuration, $\xi=\delta/\sqrt{2}$. The resulting dispersion is shown in Fig. 4a-b and demonstrates excitations dispersing from a wavevector near (1/2,1/2) ($\xi$=0.5) towards (1,0) ($\xi$=0). The shape of the dispersion is reminiscent of the "hour-glass" dispersion observed in the cuprates[25,29,30]. However, unlike the cuprates, the high energy excitations of $Fe_{1+y}Te_{1-x}Se_x$ remain centered near (1,0) with no evidence for dispersion away from (1,0). The difference between the two concentrations is emphasized in Fig. 4c. For energies greater than about 50 meV, the dispersions are consistent for the two concentrations. However, for lower energies, $|\xi|$ is larger for x=0.49 indicating excitations displaced closer to the (1/2,1/2) wavevector ($\xi$=0.5) with larger Se content. This can also be clearly seen in comparing the cuts in Figs. 4d with 4f (25 meV) and 4e with 4g (70 meV). The peaks in the 25 meV data are at clearly different wavevectors, but the distribution of scattering intensity is nearly identical at 70 meV for the two concentrations. It is interesting to note that the sample (x=0.49) with normal state excitations closer to the resonance wavevector exhibits bulk superconductivity while only weak superconductivity is observed in the sample with normal state excitations displaced further from this wavevector. However, attributing such a wavevector shift to enhanced superconductivity is complicated by the presence of excess Fe in the x=0.27 sample.

Finally, we comment on the nature of the magnetic excitations in these materials. The itinerant or local moment nature of the magnetism in Fe-based superconductors is an important, unresolved question. Local moment antiferromagnetic correlations, i.e. spin waves, would result in rings of scattering in constant energy slices whose diameter increases with increasing energy transfer. The observed spectrum (Figure 2) consists of incommensurate peaks at low energy that disperse toward the (1,0) wavevector inconsistent with expectation for local moment spin waves. Another explanation for the quartet of peaks around ($\pi$,$\pi$) in the cuprates was overlapping excitations from a stripe model[30]. In this scenario, streaks of intensity produce enhancement at intersecting wavevectors. The large value of incommensuration in $Fe_{1+y}Te_{1-x}Se_x$ makes such an explanation unlikely. To generate the quartet of peaks around (1,0) requires streaks propagating along H and K. However, such streaks would also generate considerable overlap between neighboring zones as the incommensuration is large. This is best shown in Fig. 2b where overlapping stripes along H and K would result in considerable intensity near (-0.7,-0.7), clearly not present in the data. Ruling out local moment antiferromagnetism and stripes, it is reasonable to conclude that the observed excitations for $Fe_{1+y}Te_{1-x}Se_x$ are predominately itinerant in nature.

1. Kamihara, Y., *et. al.* Iron-based layered superconductor $La[O_{1-x}F_x]FeAs$ (x = 0.05-0.12) with $T_C$ = 26 K. J. Am. Chem. Soc. **130**, 3296-3297 (2008).

2. Chen, X. C., *et al.* Superconductivity at 43 K in $SmFeAsO_{1-x}F_x$. Nature **453**, 761-762 (2008).

3. Rotter, M., Tegel & M., Johrendt, D. Superconductivity at 38 K in the iron arsenide $(Ba_{1-x}K_x)Fe_2As_2$. Phys. Rev. Lett. **101**, 107006 (2008).

4. Christianson, A. D., *et al.* Unconventional superconductivity in $Ba_{0.6}K_{0.4}Fe_2As_2$ from inelastic neutron scattering. Nature **456**, 930-932 (2008).

5. Lumsden, M. D., *et al.* Two-dimensional resonant magnetic excitation in $BaFe_{1.84}Co_{0.16}As_2$. Phys. Rev. Lett **102**, 107005 (2009).

6. Chi, S., *et al.* Inelastic neutron-scattering measurements of a three-dimensional spin resonance in the FeAs-based $BaFe_{1.9}Ni_{0.1}As_2$ superconductor. Phys. Rev. Lett. **102**, 107006 (2009).

7. Mook, H. A., *et al.* Neutron scattering patterns show superconductivity in $FeTe_{0.5}Se_{0.5}$ likely results from itinerant electron fluctuations. arXiv:0904.2178 (2009).

8. Cheong, S-W., *et al.*, Incommensurate magnetic fluctuations in $La_{2-x}Sr_xCuO_4$. Phys. Rev. Lett. **67**, 1791-1794 (1991).

9. Mason, T.E., et al., Magnetic dynamics of superconducting $La_{1.86}Sr_{0.14}CuO_4$. Phys. Rev. Lett. **68**, 1414-1417 (1992).

10. Dai, P., Mook, H. A., and Dogan, F., Incommensurate magnetic fluctuations in $YBa_2Cu_3O_{6.6}$, Phys. Rev. Lett. **80,** 1738-1741 (1998).

11. Mook, H. A. et al., Spin fluctuations in YBa2Cu3O6.6, Nature **395**, 580-582 (1998).

12. Chen, G. F., et al., Superconductivity at 41 K and Its Competition with Spin-Density-Wave Instability in Layered $CeO_{1-x}F_xFeAs$. Phys. Rev. Lett. **100** 247002 (2008).

13. Ren, Z. A., *et al.* , Superconductivity in the iron-based F-doped layered quaternary compound $NdO_{1-x}F_x$] FeAs. Europhys. Lett. **82** 57002 (2008).

14. Ren Z, *et al.*, Superconductivity at 55 K in Iron-Based F-Doped Layered Quaternary Compound Sm[$O_{1-x}F_x$] FeAs, Chin. Phys. Lett. **25,** 2215.

15. Wang, X. *et al.*, The superconductivity at 18 K in LiFeAs system. Solid State Comm. 148, 538 (2008).

16. Hsu, F.-C. *et al.*, Superconductivity in the PbO-type structure α-FeSe. PNAS 105, 14262 (2008).

17. Yeh, K.-W. *et al.*, Tellurium substitution effect on superconductivity of the α-phase iron selenide. Europhys Lett. **84** (2008) 37002.

18. de la Cruz, C. *et al.*, Magnetic order close to superconductivity in the iron-based layered $LaO_{1-x}F_xFeAs$ systems. Nature **453**, 899 (2008).

19. McGuire, M. A. *et al.*, Phase transitions in LaFeAsO: Structural, magnetic, elastic, and transport properties, heat capacity and Mössbauer spectra. Phys. Rev. B **78**, 094517 (2008).

20. Huang, Q. *et al.*, Neutron-Diffraction Measurements of Magnetic Order and a Structural Transition in the Parent $BaFe_2As_2$ Compound of FeAs-Based High-Temperature Superconductors. Phys. Rev. Lett. **101**, 257003 (2008).

21. Bao, Wei, et al., Tunable (δπ, δπ)-Type Antiferromagnetic Order in α-Fe(Te,Se) Superconductors. Phys. Rev. Lett. **102**, 247001 (2009).

22. Qiu, Y., *et al.*, Spin Gap and Resonance at the Nesting Wavevector in Superconducting $FeSe_{0.4}Te_{0.6}$. arXiv:0905.3599 (2009).

23. Sales, B. C., *et al.*, Bulk superconductivity at 14 K in single crystals of $Fe_{1+y}Te_xSe_{1-x}$. Phys. Rev. B 79, 094521 (2009).

24. Stock, C., *et al.*, From incommensurate to dispersive spin fluctuations: The high energy inelastic spectrum in superconducting $YBa_2Cu_3O_{6.5}$. Phys. Rev. B **71,** 024522 (2005).

25. Vignolle, B., *et al.*, Two energy scales in the spin excitations of the high-temperature superconductor $La_{2-x}Sr_xCuO_4$. Nature Physics **3**, 163 (2007).

26. Wen, J., *et al.*, Coexistence and competition of short-range magnetic order and superconductivity in $Fe_{1+\delta}Te_{1-x}Se_x$, arXiv: 0906.3774 (2009).

27. Zhang, L., Singh, D. J. & Du, M. H. Density functional study of excess Fe in $Fe_{1+x}Te$: Magnetism and doping. Phys. Rev. B **79**, 012506 (2009).

28. Sato, H. and Maki, K., Theory of inelastic neutron scattering from Cr and its alloys near the Néel temperature. Int. J. Magnetism **6**, 183 (1974).

29. Hayden, S. M. *et al.*, The structure of the high-energy spin excitations in a high-transition-temperature superconductor. Nature **429**, 531 (2004).

30. Tranquada, J.M. *et al.*, Quantum magnetic excitations from stripes in copper oxide superconductors. Nature **429**, 534 (2004).

Author Contributions Statement: All authors made critical comments on the manuscript. M. L, A. C., E. G., S. N., M. S., D. A., T. G., G. M., C. C., and H. M. all contributed to data collection. A. S. M. M., B. S., and D. M. contributed to sample synthesis and characterization.

We acknowledge discussions with David Singh. This work was supported by the Scientific User Facilities Division and the Division of Materials Sciences and Engineering, Office of Basic Energy Sciences, US Department of Energy. We acknowledge discussions with David Singh.

Correspondence and requests for materials should be addressed to M. L. (lumsdenmd@ornl.gov).

Figure 1: (a) Reciprocal space for $Fe_{1+y}Te_{1-x}Se_x$ compounds. Black (red) labels correspond to wavevectors in the tetragonal (square) reciprocal lattice. Red

circles represent the square lattice (π,0) points where the resonance is observed while green squares represent the square lattice (π,π) points. (b) The same reciprocal space diagram as (a) but centred at (1, 0). The blue ovals show the location of the observed incommensurate excitations and the arrows represent the direction of dispersion with increasing energy transfer.

Figure 2: Constant energy plots of the magnetic excitations in $Fe_{1.04}Te_{0.73}Se_{0.27}$, a-d, and $FeTe_{0.51}Se_{0.49}$, e-h. All measurements were performed with the c-axis parallel to the incident beam and the sample temperature was 5 (3.5) K for the x=0.27 (0.49) measurements. Panels a-d show data from the x=0.27 sample measured using MERLIN with incident energies of 25, 60, 120, and 250 meV, respectively. Panels e-h show data from the x=0.49 sample measured using ARCS with incident energies of 40, 60, 120, and 250 meV, respectively. The x=0.27(0.49) samples were single crystals with a mass of 16.91(15.45) g. The data is fit with a model function (equation 1) convolved with instrumental resolution and the resulting fits are shown in i-l (m-p) for the x=0.27(0.49) sample. The fit (i) to the 10 meV data (a) for the x=0.27 sample includes two components rotated by 45 degrees in the H-K plane. All other fits include only a single component. All plots show a projection of the data onto the H-K plane.

Figure 3: Constant energy plots through the magnetic excitation spectrum at an energy transfer of 6±1 meV for both the x=0.27 and x=0.49 samples. Measurements of the x=0.27 (0.49) sample a, (b) were performed with an incident energy of 25 (40) meV. The temperature was 5 (3.5) K for the x=0.27 (0.49) measurements. This shows an additional component in the x=0.27 data centred near (1/2, 0) which is absent in the x=0.49 data. The fits to a model function (equation 1) convolved with instrumental resolution (c-d). Fits to the x=0.27 data (c) included two components rotated by 45 degrees with respect to one another in the H-K plane while fits to the x=0.49 data (d) included only the single, incommensurate component.

Figure 4: Dispersion of the magnetic excitation spectrum in $Fe_{1+y}Te_xSe_{1-x}$ for x=0.27 (a) and x=0.49 (b). The dispersion, $\xi$, is related to $\delta$ (see equation 1) as $\xi=\delta/\sqrt{2}$. Below 15 meV, the dispersion was extracted from triple-axis measurements in the (HK0) scattering plane via a convolution of equation 1 with instrumental resolution. The dispersion above 15 meV was extracted from the time-of-flight data. For the x=0.27 sample, all data analyzed was measured at 3.5 K (triple-axis) or 5 K (time-of-flight). For the x=0.49 sample, time-of-flight data analyzed was measured at 3.5 K while the triple-axis data analyzed was measured at 20 K to avoid scattering from the magnetic resonance. Fits to the time-of-flight data were performed both with and without inclusion of instrumental resolution. As resolution was found to have very little effect on the extracted value of $\delta$, the dispersion was generated without the inclusion of resolution for simplicity. The fits generate a single value of $\delta$ for each energy transfer. The plotted dispersion is symmetrized to illustrate the symmetry about the (1, 0) wavevector. For energies above 100 meV, the peak position cannot be distinguished from $\xi$=0. The inset, (c), shows a direct comparison of the dispersion for the two concentrations demonstrating low energy excitations closer to (1/2,1/2) for the x=0.49 sample. Cuts at 25±3 meV were measured with incident energy of 120 meV for the x=0.27 (d) and x=0.49 (f) samples. Cuts at 70±5 meV were measured with incident energy of 250 meV for the x=0.27 (e) and x=0.49 (g) samples. Solid lines represent fits using equation 1 showing excellent agreement with the data.

Figure 1:

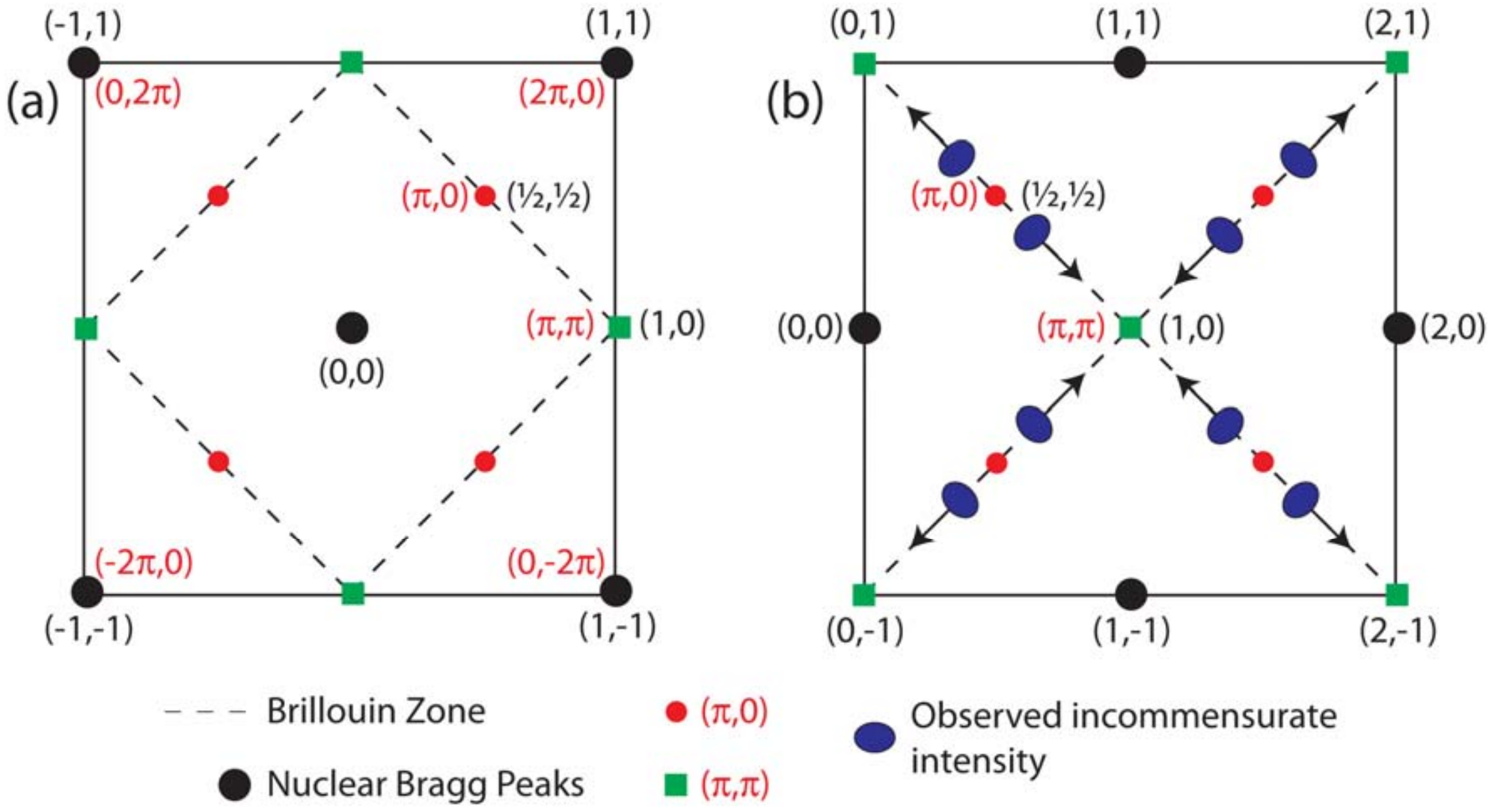

Figure 2.

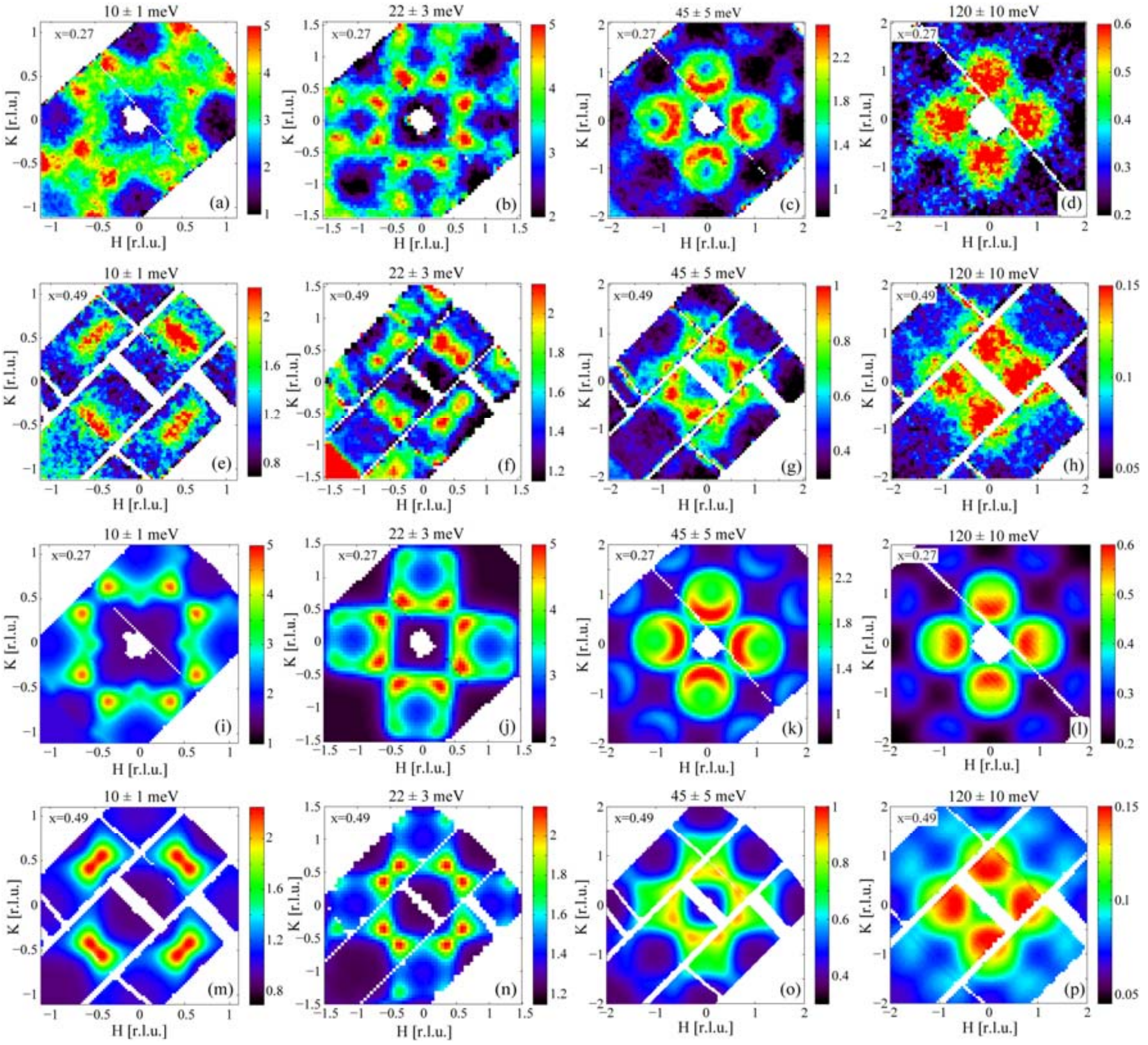

Figure 3:

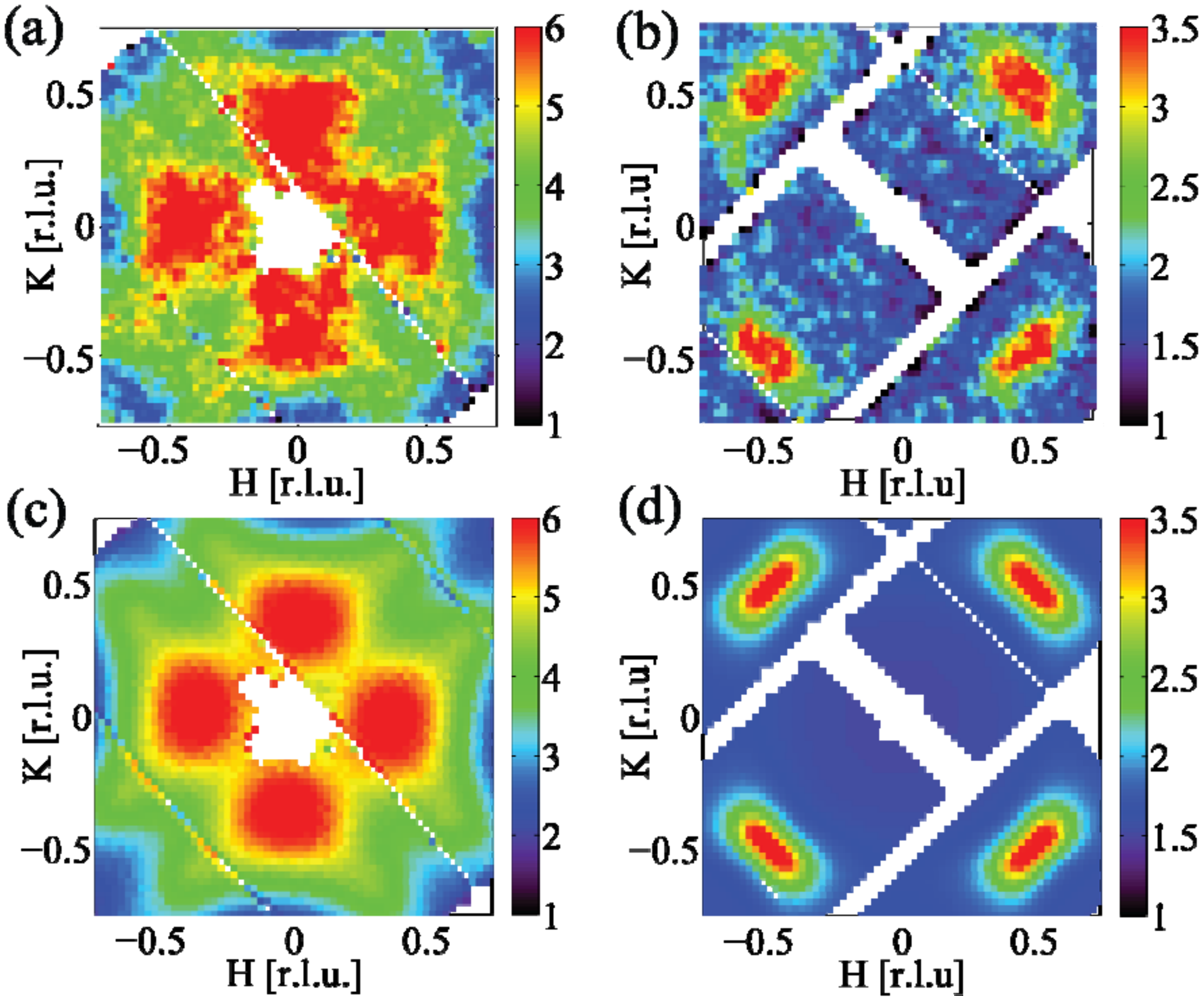

Figure 4:

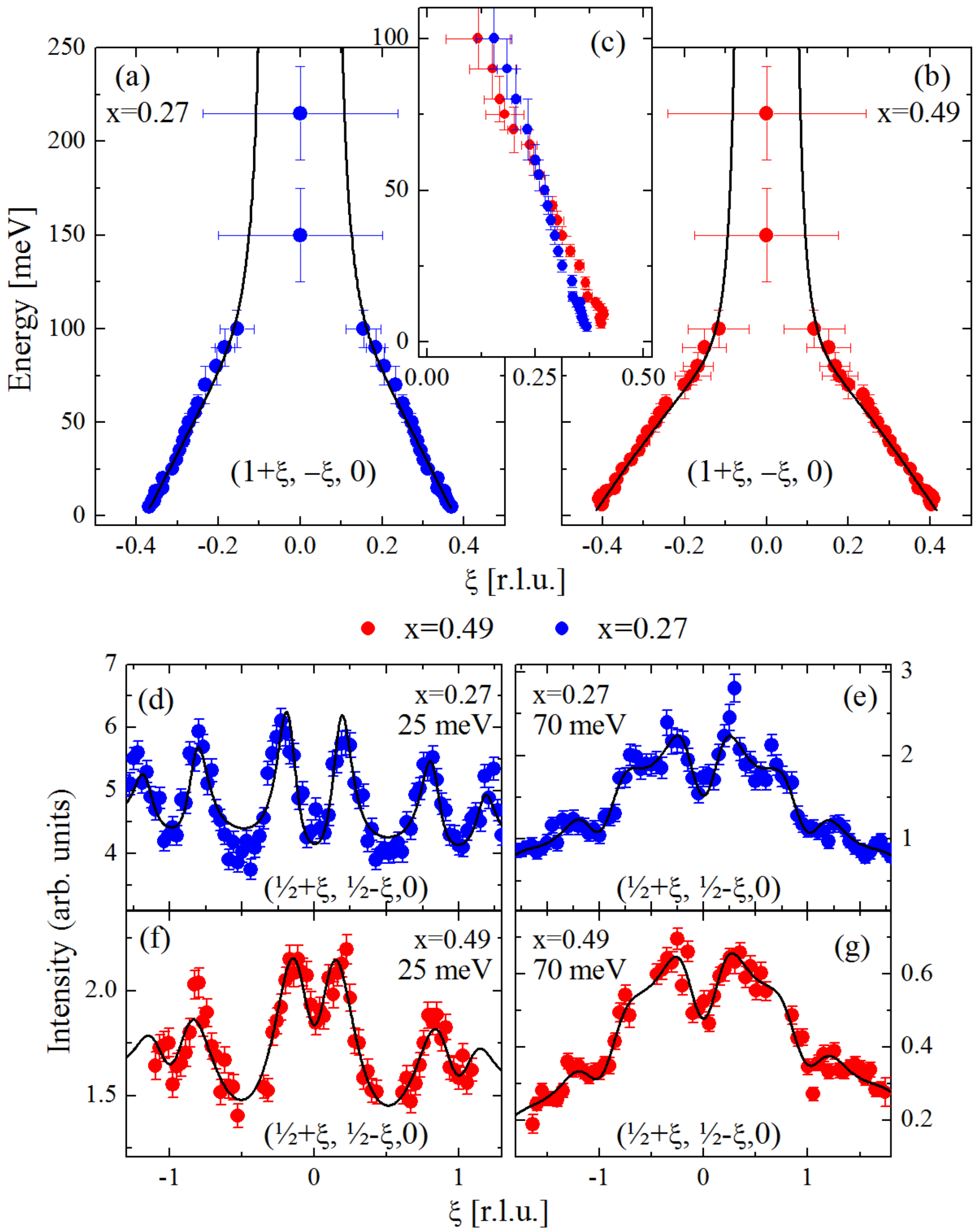

# Supplementary information

The lack of L dependence in the measurements can be demonstrated by collecting data with the sample rotated by 90 degrees from those shown in Fig. 2. In this geometry, the (1 1 0) direction is along the incident beam. The figure below shows the data obtained in this geometry for energies of 22±3 meV and 45±5 meV to allow for direct comparison with Fig. 2 (b) and (c). One can immediately see that the intensity extends over a range of L values much wider than a Brillouin zone suggesting weak L dependence. Quantitative understanding of the extent along L is complicated by the fact that the data has been integrated over the component of Q along (1, 1, 0) thereby projecting the data onto the plane defined by (H, -H, 0) and (0, 0, L). To help understand this, fits (panels (b) and (d)) are obtained using equation 1 with the identical values of $\kappa$, $\delta$, and $\gamma$ as those presented in Fig. 2(b) and 2(c). It is important to note that the model used (equation 1) is entirely two-dimensional. One can clearly see that the fits reproduce the extent of the data along L for both energy transfers showing that a 2d model provides an excellent description of the data. We take this as very strong evidence that the excitations in this material are two-dimensional.

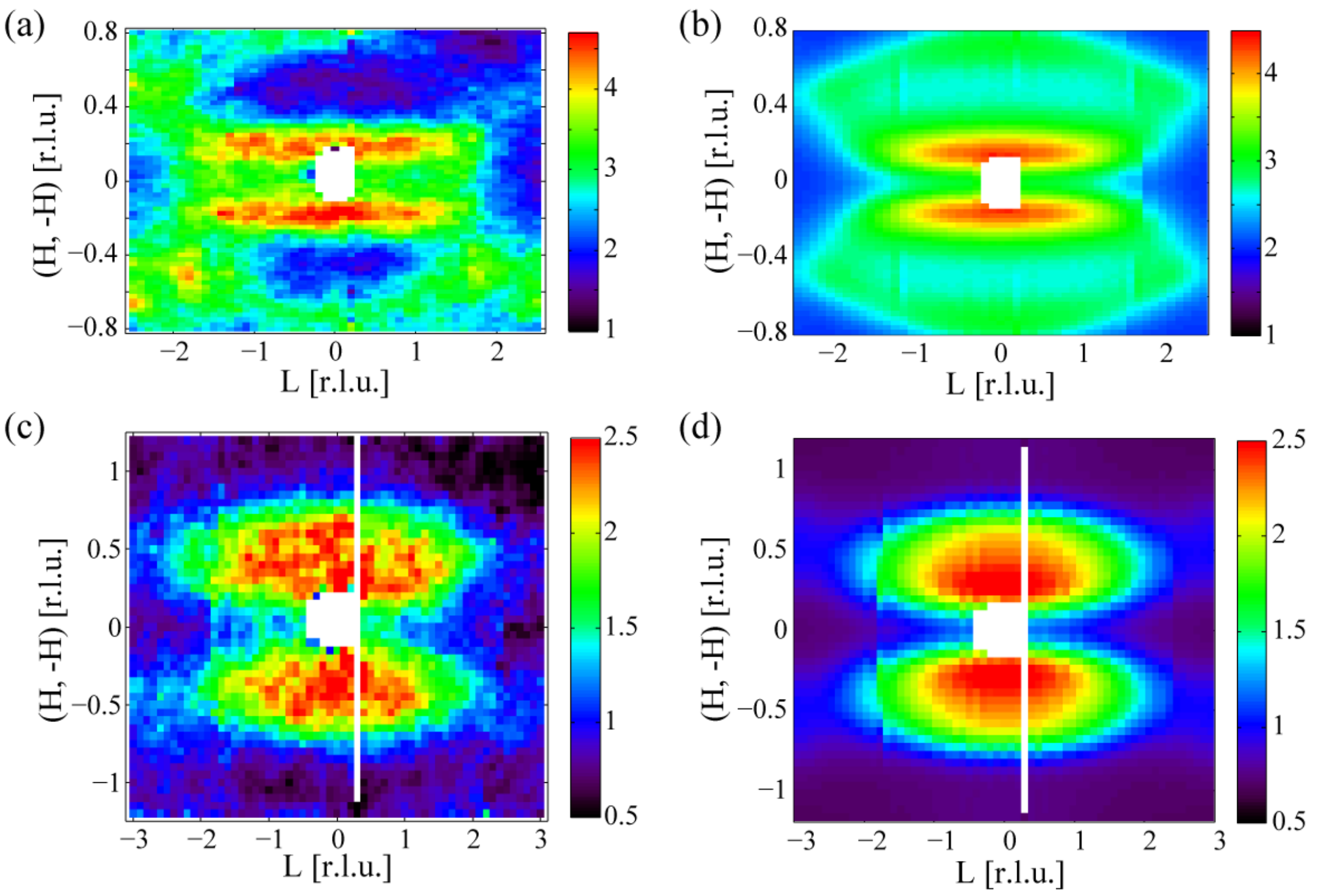

Figure 1. L-dependence of the spin excitations in $Fe_{1.04}Te_{0.73}Se_{0.27}$. Measurements were performed with the (110) axis parallel to the incident beam with data projected onto the plane defined by (H, -H, 0) and (0, 0, L). (a) and (c) show constant energy plots at energies of 22±3 and 45±5 meV respectively. These energies are the same as those in Fig. 2(b) and 2(c). The fits in (b) and (d) are obtained using equation 1 with values of κ, δ, and γ set to be identical as those obtained with the c-axis parallel to the incident beam.

The expression for the Sato-Maki function (equation 1) can be generalized to describe both the incommensurate excitations and the component at (1/2,0) (x=0.27 sample only). If we define ϕ as a general rotation angle for the scattering pattern, we can define,

$$\mathbf{q}_1 = \mathrm{H}\cos\phi + \mathrm{K}\sin\phi \qquad \mathbf{q}_2 = -\mathrm{H}\sin\phi + \mathrm{K}\cos\phi$$

$$\mathbf{q_{1C}} = \mathrm{H_C}\cos\phi + \mathrm{K_C}\sin\phi \qquad \mathbf{q_{2C}} = -\mathrm{H_C}\sin\phi + \mathrm{K_C}\cos\phi . \tag{1}$$

If we further define

$$\mathbf{r_1} = (\mathbf{q_1} - \mathbf{q_{1C}})^2 \text{ and } \mathbf{r}_2 = (\mathbf{q}_2 - \mathbf{q}_{2C})^2 , \tag{2}$$

we can write a generalized version of R(Q) as

$$R(\mathbf{Q}) = \frac{(\mathbf{r_1}+\mathbf{r_2}-\delta^2)^2 + \lambda \mathbf{r_1}\mathbf{r_2}}{4\delta^2} . \tag{3}$$

The above expression for R(**Q**) can reproduce the expression (equation 2) describing the incommensurate excitation by setting $\phi=\pi/4$ and can generate an expression describing the data near (1/2,0) by setting $\phi=0$.